\documentclass[aps,prl,twocolumn,groupaddress,nofootinbib,longbibliography]{revtex4-2}
\usepackage{graphicx}
\usepackage{amsmath}
\usepackage{subfigure}
\usepackage{braket}
\usepackage{color}

\newcommand*{\doublebar}[1]{\overline{\overline{#1}}}

\begin{document}

\title{Effects of higher-order Casimir-Polder interactions on Rydberg atom spectroscopy}

\author{B. Dutta}
\affiliation{Laboratoire de Physique des Lasers, Universit{\'e} Sorbonne Paris Nord, F-93430, Villetaneuse, France}
\affiliation{CNRS, UMR 7538, LPL, 99 Avenue J.-B. Cl{\'e}ment, F-93430 Villetaneuse, France} 

\author{J. C. de Aquino Carvalho}
\affiliation{Laboratoire de Physique des Lasers, Universit{\'e} Sorbonne Paris Nord, F-93430, Villetaneuse, France}
\affiliation{CNRS, UMR 7538, LPL, 99 Avenue J.-B. Cl{\'e}ment, F-93430 Villetaneuse, France}

\author{G. Garcia-Arellano}
\affiliation{Laboratoire de Physique des Lasers, Universit{\'e} Sorbonne Paris Nord, F-93430, Villetaneuse, France}
\affiliation{CNRS, UMR 7538, LPL, 99 Avenue J.-B. Cl{\'e}ment, F-93430 Villetaneuse, France}  

\author{P. Pedri}
\affiliation{Laboratoire de Physique des Lasers, Universit{\'e} Sorbonne Paris Nord, F-93430, Villetaneuse, France}
\affiliation{CNRS, UMR 7538, LPL, 99 Avenue J.-B. Cl{\'e}ment, F-93430 Villetaneuse, France} 

\author{A. Laliotis}
\email{laliotis@univ-paris13.fr}
\affiliation{Laboratoire de Physique des Lasers, Universit{\'e} Sorbonne Paris Nord, F-93430, Villetaneuse, France}
\affiliation{CNRS, UMR 7538, LPL, 99 Avenue J.-B. Cl{\'e}ment, F-93430 Villetaneuse, France} 

\author{C. Boldt}
\affiliation{Institute for Physics, University of Rostock, Albert-Einstein-Stra{\ss}e 23-24, D-18059 Rostock, Germany}

\author{J. Kaushal}
\affiliation{Institute for Physics, University of Rostock, Albert-Einstein-Stra{\ss}e 23-24, D-18059 Rostock, Germany}

\author{S. Scheel}
\affiliation{Institute for Physics, University of Rostock, Albert-Einstein-Stra{\ss}e 23-24, D-18059 Rostock, Germany}

\date{\today}

\begin{abstract}
 
In the extreme near-field, when the spatial extension of the atomic wavefunction is no longer negligible 
compared to the atom-surface distance, the dipole approximation is no longer sufficient to describe 
Casimir-Polder interactions. Here we calculate the higher-order, quadrupole and octupole, contributions 
to Casimir-Polder energy shifts of Rydberg atoms close to a dielectric surface. We subsequently 
investigate the effects of these higher-order terms in thin-cell and selective reflection spectroscopy. 
Beyond its fundamental interest, this new regime of extremely small atom surface separations is relevant 
for quantum technology applications with Rydberg or surface-bound atoms interfacing with photonic 
platforms.

\end{abstract}


\maketitle

Highly excited (Rydberg) atoms have huge electric and magnetic transition multipole moments that make 
them interact very strongly with their 
environment. Thus, they are ideal candidates for studying dispersion forces such as 
Casimir-Polder (atom-surface) \cite{SandoghdarPRL1992, SandoghdarPRA1996} or van der Waals (atom-atom) 
interactions \cite{BrowaeysPRL2013}. More recently, Rydberg atoms have attracted significant 
attention for quantum technology applications. In particular, it was demonstrated that probing Rydberg
atoms inside thin vapour cells \cite{kublernatphot2010} presents a simple way to fabricate room 
temperature single-photon sources based on the Rydberg blockade effect \cite{ripkascience2018}, without 
having to resort to complex cold atom manipulations. Moreover, Rydberg atoms in vapour cells are being 
used as electric-field probes at THz or GHz frequencies \cite{Sedlacek2012,wadenatphot2017, DownesPRX2020}. However, the rapid scaling of electric-dipole moment fluctuations 
makes Casimir-Polder (CP) interactions a dominant spectroscopic contribution that impacts potential 
hybrid systems such as tapered optical fibers \cite{RajasreePRR2020, vylegzhanin&Chormaic_2023}, hollow core fibers 
\cite{epplenatcom2014} and thin cells \cite{kublernatphot2010, ripkascience2018} that aim at 
interfacing atoms with photonic platforms. Similarly, Rydberg interactions with van der Waals 
heterostructures have also been investigated \cite{wongcharoenbhornPRA2023}.  

Theoretical studies of the interaction between dielectric surfaces and highly excited atoms have exposed
the limitations of the traditional Casimir-Polder (CP) approach in which only the electric dipole 
interaction is taken into account. Indeed, the dipole approximation breaks down and higher-order terms 
need to be considered in the extreme near field \cite{crossePRA2010} as Rydberg wavefunctions can easily
extend beyond 100nm (being proportional to ${n^\star}^2$, with ${n^\star}$ the effective quantum 
number). Moreover, perturbative approaches have also been questioned when the expected energy shifts are 
comparable to adjacent energy level spacings \cite{RibeiroEPL2015}. Although previous theoretical works 
have presented the basic scaling laws governing quadrupole interactions \cite{crossePRA2010}, detailed results of the higher order interaction coefficients have not yet been presented for CP atom-surface interactions. Higher order effects have nevertheless been studied in detail for the case of atom-atom or molecule-molecule van der Waals interactions \cite{Salam1994, Salam1996}.

Atom-surface experiments have similarly flourished during the past decades \cite{Bender_PRL_2010, Carvalho_PRL_2023, cornell_PRL_2007, cornell_PRA_2004}, shedding light on novel effects such as the 
temperature dependence of CP interactions \cite{cornell_PRL_2007, laliotisnatcommun2014}, and atom-
metamaterial interactions \cite{laliotis_scienceadv2018} with an ever increasing precision in view of 
uncovering fundamental forces beyond the standard model \cite{cornell_PRA_2004}. However, higher-order 
interactions remain so far experimentally unexplored. A prominent and well developed experimental method 
for performing such studies is vapour cell spectroscopy
\cite{oriaepl1991, failacheprl1999, laliotisnatcommun2014, fichet_epl_2007, peyrotpra2019}. Thin vapour
cells allow for a controlled confinement of atomic vapours within dielectric walls down to the nanometer 
regime, making them excellent platforms for probing Rydberg atoms extremely close to dielectric surfaces 
\cite{kublernatphot2010, fichet_epl_2007}. Selective reflection in macroscopic cells has also been used 
for atomic or molecular gases \cite{oriaepl1991,laliotisnatcommun2014, lukusa_prl_2021} close to a surface, but provides no means of controlling the probing 
depth, typically defined by the excitation wavelength. 

Here, we investigate higher-order (quadrupole and octupole) CP interactions between a Rydberg atom and a 
dielectric surface, providing explicit calculations of the $C_3$ (dipole-dipole) and the $C_5$ (combined 
quadrupole-quadrupole and dipole-octupole) coefficients for alkali Rydberg atoms in the near field. We 
subsequently study the effects of higher-order interactions on CP vapour cell spectroscopy, 
illuminating the conditions under which higher-order interactions can be experimentally measurable. 
Multipole contributions are of importance for experiments aiming at binding atoms to surfaces and 
extending the optical control to the extreme near field \cite{HummerPRL2021}.     

We will conduct our studies in the electrostatic limit, where CP Rydberg-surface interactions can be 
described as the interaction of the atomic charge distribution (centred in $O$) with its surface-induced instantaneous 
image (centred in $O^\prime$) in front of a perfectly conducting surface. We assume that the Rydberg atom consists of a valence 
electron orbiting around a positively charged core. The atom-surface interaction energy, $W$, is half 
the electrostatic energy of the atomic charge distribution placed under the external potential of its 
image, $\Phi^{\prime}(\vec{r^\prime})$, with the corresponding field
$\vec{E^{\prime}}(\vec{r^\prime})$,
\begin{equation}
\begin{split}
    W=&-\frac{1}{2}\vec{p}\cdot \vec{E^{\prime}}(\vec{r}_0)-\frac{1}{12}\sum_{i,j} Q_{ij} \frac{\partial {E^{\prime}_i} }{\partial {r_j^\prime}} (\vec{r}_0)+\\
    &-\frac{1}{24}\sum_{i,j,k} T_{ijk} \frac{\partial^2 {E^{\prime}_i} }{\partial{r_j^\prime} \partial {r_k^\prime}} (\vec{r}_0) +\dots
\end{split}
\end{equation}
Here, $\vec{r}_0$ is the position vector of the atom 
with coordinates ${r_i}$. 
The atomic multipole moments with respect to $O$ are denoted as $\vec{p}$, $\bar{Q}$ and $\doublebar{T}$ 
(dipole, quadrupole and octupole moments, respectively). The potential created by the image can itself 
be expanded into multipoles, giving
\begin{equation}
\begin{split}
\Phi^{\prime}(\vec{r}_0)=&\frac{1}{4\pi\epsilon_0}\sum_i{p^\prime_i} \frac{{r}_{0i}}{r_0^3}+\frac{1}{8\pi\epsilon_0}\sum_{i,j}{Q^\prime_{ij}} \frac{r_{0i} r_{0j}}{r_0^5}+\\
&+\frac{1}{24\pi\epsilon_0}\sum_{i,j,k} T^\prime_{ijk} \frac{r_{0i} r_{0j} r_{0k}}{r_0^7}+\dots
\end{split}
\end{equation}
Here, $\vec{p^\prime}$, $\bar{Q^\prime}$, and $\doublebar{T^\prime}$ are the dipole, quadrupole and 
octupole moments of the image with respect to $O^\prime$. Symmetry links the components of the 
image moments to those of the atomic moments by  $p_i^\prime=(-1)^{\kappa + 1} p_i$, 
$Q_{ij}^\prime=(-1)^{\kappa +1}Q_{ij}$, $T_{ijk}^\prime=(-1)^{\kappa +1}T_{ijk}$, 
where $\kappa$ is the number of times $z$ appears as a tensor index. 

The CP interaction energy can therefore be written as
\begin{equation}
    W = W_{pp}+W_{pQ}+W_{QQ}+W_{pT}.
\end{equation}
where $W_{pp}$, $W_{pQ}$, $W_{QQ}$, $W_{pT}$  are the dipole-dipole, dipole-quadrupole, quadrupole-
quadrupole and dipole-octupole contributions, respectively. The dipole-dipole and quadrupole-quadrupole 
terms are given by
\begin{gather}
W_{pp} = - \frac{1}{4\pi\epsilon_0}\frac{p^2 + p_{z}^2}{16z_s^3}, \\
W_{QQ} = -\frac{17Q_{zz}^2 + 16Q_{zy}^2 + 16Q_{zx}^{2} + 2Q_{xx}^{2} + 4Q_{yx}^{2} + 2Q_{yy}^{2}}{4\pi\epsilon_0 768 z_s^5}  .
\end{gather}
The cross terms such as the dipole-quadrupole and dipole-octupole contributions with their $z_s^{-4}$ 
and $z_s^{-5}$ dependence, respectively, are given in the Supplementary material. It should be noted 
that $W_{pQ}$ includes both the energy of a dipole immersed in the field of a quadrupole and vice-versa 
(the same applies to $W_{pT}$).

In the electrostatic limit, we can calculate the CP frequency shift, $\Delta f$, of an atomic Rydberg 
state using first order perturbation theory as
\begin{equation}
h \Delta f=\langle \psi_{n,l,J,M_{J}}\vert W \vert \psi_{n,l,J,M_{J}} \rangle  \,.
\end{equation}
Here, $\psi_{n,l,J,M_{J}}$ is the wavefunction of the Rydberg electron, with $n$ the principal quantum 
number, $l$, $J$ the orbital and total angular momentum quantum numbers, and $M_J$ the projection of $J$ 
onto the z-axis. For Rydberg atoms we can assume that the external electron is under the influence of a 
central potential given by an effective Coulomb interaction modified to include the polarisability of 
core electrons \cite{Derevianko99}. This allows us to easily obtain the radial wavefunction by 
numerically solving Schr\"odinger's equation using the Numerov method.  
In our analysis, we ignore the hyperfine structure of the atoms as it is usually very small compared to 
the CP energy shifts for most alkali Rydberg atoms. The above analysis allows us to calculate Rydberg-surface interactions at any multipole order.

\begin{figure}[!t]
\includegraphics[width=85mm]{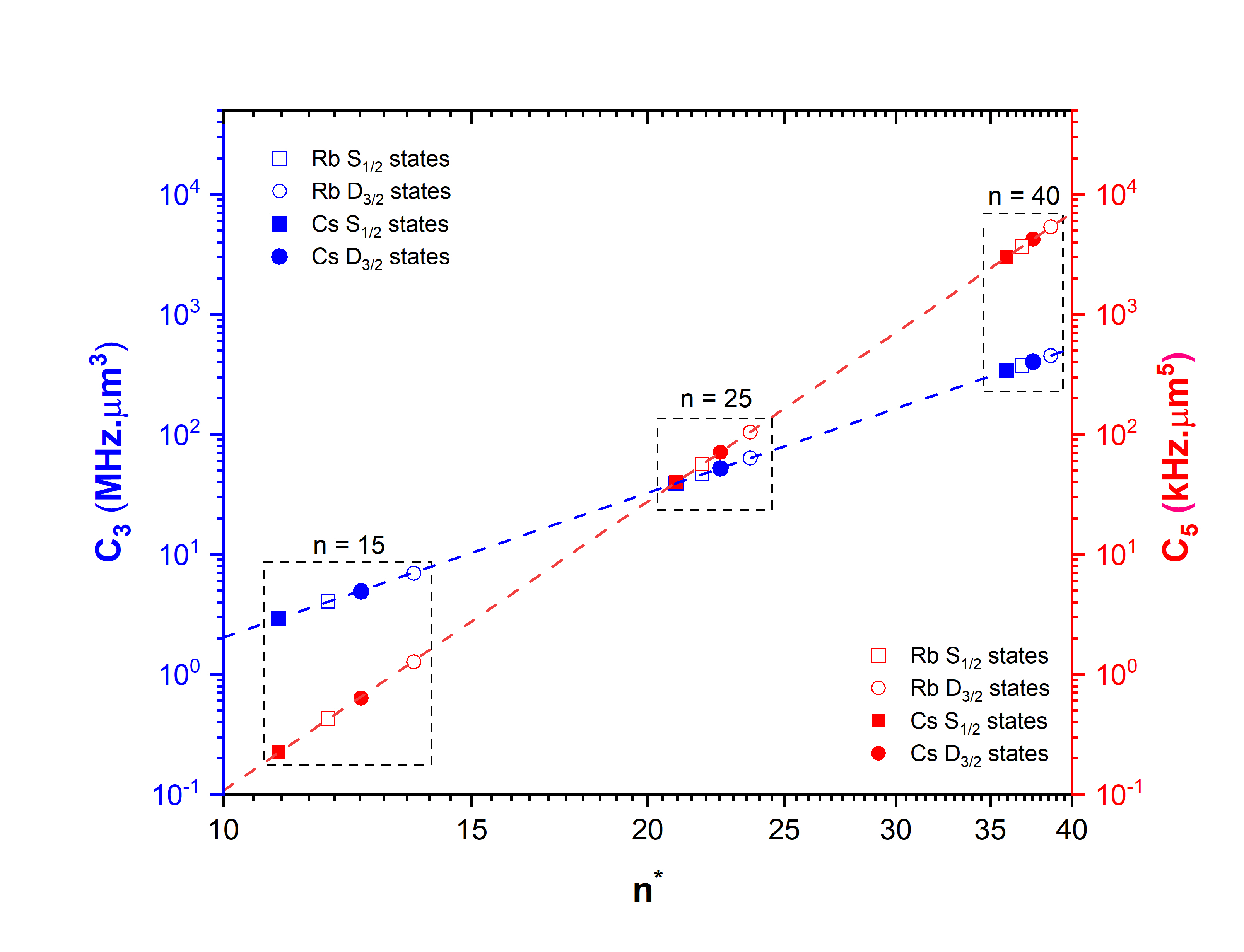}%
\caption{$C_3$ and $C_5$ coefficients (blue and red colors, respectively) for $S_{1/2}$ and $D_{3/2}$ 
states (squares and circles, respectively) of cesium and rubidium (open and closed points, respectively) 
as a function of the effective quantum number $n^{\star}=n-\delta$. The straight solid lines correspond 
to the analytical expressions derived for a hydrogen $S$ state [Eqs.~(\ref{eq:C3}) and (\ref{eq:C5})] 
providing an accurate estimate, to within 1$\%$, of the interaction coefficients for all alkali Rydberg 
atoms.  
\label{Fig1}}
\end{figure}

The quantum mechanical averaging of the interaction energy gives the following results: a) The dipole-dipole term ($\Delta f_{3}$) yields the well-known near-field Casimir-Polder frequency shift 
$\Delta f_{3}=-C_3/z_s^3$, where $C_3$ is a coefficient usually expressed in $\mathrm{MHz\,\mu m^3}$. 
b) The dipole-quadrupole terms vanish for parity reasons.
c) The quadrupole-quadrupole and dipole-octupole terms give a frequency shift 
expressed as $\Delta f_{5}=-C_5/z_s^5$, where $C_5$ is a coefficient expressed in 
$\mathrm{MHz\,\mu m^5}$. We emphasize that the dipole-octupole contributions do not necessarily vanish 
as both dipole and octupole interactions can act on the same atomic transition ($\Delta l=\pm 1$ 
transitions can be both dipole and octupole allowed). 
However, dipole-octupole terms only contribute to the anisotropy of the atom-surface interaction, with 
an overall scalar component (averaging over all $M_J$) that remains zero. 

In Fig.~\ref{Fig1} we plot the $C_3$ and $C_5$ coefficients for $S_{1/2}$ and $D_{3/2}$ states of cesium and rubidium 
atoms as a function of the effective quantum number ($n^*=n-\delta^*$), where $\delta^*$ is the quantum 
defect. It shows that the interaction coefficients depend very little on the 
core polarisability and on angular momentum. We have therefore 
calculated analytical expressions for both $C_3$ and $C_5$ coefficients for a hydrogen atom. For an 
$S_{1/2}$ state ($l=0$) the interaction coefficients become
\begin{gather}
\label{eq:C3}
C_3=\frac{e^2 a_o^2}{96\pi\epsilon_0 h}{n^\star}^2\left(5{n^\star}^2+1\right) \approx \frac{5 e^2 a_o^2 {n^\star}^4}{96\pi\epsilon_0 h}, \\
\label{eq:C5}
C_5=\frac{e^2 a_o^4}{640\pi\epsilon_0 h}{n^\star}^4\left(63{n^\star}^4+105{n^\star}^2+ 12\right)\approx \frac{63 e^2 a_o^4 {n^\star}^8}{640\pi\epsilon_0 h}.
\end{gather}
The full solution, for all states, is given in the Supplemental material. Recall that for hydrogen 
${n^\star}$ is simply the principal quantum number and note that the leading term in the above 
polynomials is independent of angular momentum.

Our assumption of a perfectly correlated image breaks down at distances comparable to relevant 
transition wavelengths or when a frequency-dependent dielectric constant is considered 
\cite{fichetPRA1995}. However, for Rydberg atoms, the relevant multipole transitions 
are typically in the THz or GHz regime suggesting that 
retardation can be ignored at micrometer distances. 
Moreover, at these frequencies most materials do not possess surface resonances, and their dielectric 
constant tends to their static ($\epsilon_s$) value. Dielectric effects can therefore be accounted for 
by simply multiplying the dispersion 
coefficients (Fig.~\ref{Fig1}) by the surface response ($S=\frac{\epsilon_s-1}{\epsilon_s+1}$), while 
temperature effects \cite{gorzaepjd2006, barton1997, laliotisnatcommun2014} are negligible and can be 
ignored \cite{laliotisPRA2015}. The above arguments suggest that the electrostatic limit is a good 
approximation for studying Rydberg-surface interactions in the extreme near field. However, a QED 
treatment \cite{CasimirPR1948, wylie&SipePRA1984, wylie&SipePRA1985} would be necessary for treating the 
coupling of atoms with resonant surfaces or meta-surfaces. 

At this point, a word of caution is appropriate. Frequently, traceless multipole tensors are used in the 
calculation. However, this is only justified for quadrupole tensors, as its trace component does not 
contribute to the energy. This is no longer true for the octupole tensor, where the trace \textit{does} 
in general contribute to the energy \cite{Raab1975, RaabBook}. Indeed, upon reducing the quantity $r_ir_jr_k$ 
appearing in $\doublebar{T}$ in terms of spherical harmonics yields terms with $Y_{3m}(\Theta,\varphi)$ 
(traceless part) as well as $Y_{1m}(\Theta,\varphi)$ (trace part).
The latter can be identified as a contribution to the dipole transition, and only further 
symmetry considerations of the field distribution or atomic transition matrix elements can cause them
to vanish \cite{Salam2005, Salam2019} .

Having developed a theoretical framework that allows us to calculate both $C_3$ and $C_5$ coefficients, 
we can estimate the effects of quadrupole interactions on CP experiments. Atomic 
spectroscopy in vapour cells of variable nanometric thickness (thin cells) is a well developed method 
for measuring CP interactions with excited states including Rydberg atoms 
\cite{fichet_epl_2007, duttaep2022}. Thin cells allow us to control 
the vapour confinement down to the nanometer regime (thickness can be as small as 50nm) \cite{fichet_epl_2007}, giving them a distinct advantage over other methods for probing 
higher order CP interactions with Rydberg atoms. 

\begin{figure}[!t]
\includegraphics[width=75mm]{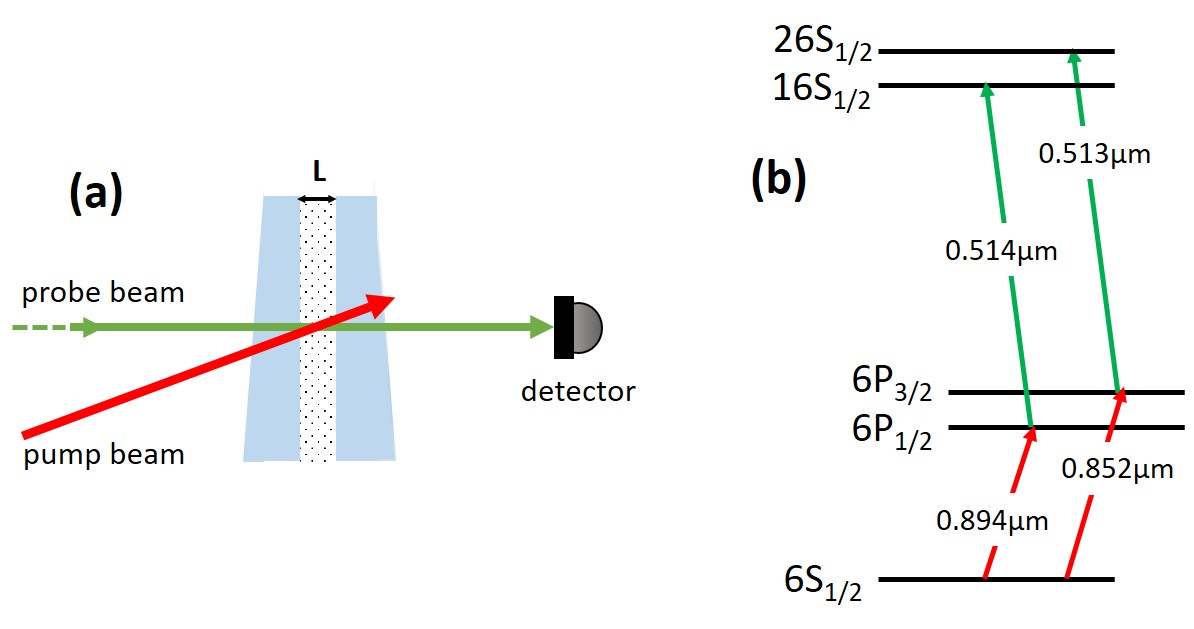}
\caption{(a) Schematic of the thin cell experiment analysed in our simulations. A laser beam at 
0.894$\mathrm{\mu m}$ or 0.852$\mathrm{\mu m}$ pumps cesium atoms to the 6$P_{1/2}$ or 6$P_{1/2}$ level,
respectively. Subsequently, a green laser at 0.514$\mathrm{\mu m}$ or 0.513$\mathrm{\mu m}$ probes
Rydberg atoms at the 6$P_{1/2}$ $\rightarrow$ 16$S_{1/2}$ or the 6$P_{3/2}$ $\rightarrow$ 26$S_{1/2}$ 
transition, respectively. Higher lying states can also be easily accessed via the same scheme. Due to interatomic collisions, the population of the intermediate levels has a 
quasi-thermal (Maxwell-Boltzmann) velocity distribution. (b) Relevant energy levels. 
\label{Fig2}}
\end{figure}

Typically, a two-step excitation technique is used to reach high-lying excited states 
of alkali atoms \cite{laliotisnatcommun2014, failacheprl1999, fichet_epl_2007}. For cesium (our atom of 
choice), a strong pump excites atoms to their first excited state (6$P_{1/2}$ or 6$P_{3/2}$) and a weak 
laser probes the 6$P_{1/2}$ $\rightarrow$  n$S$ or 6$P_{1/2}$ $\rightarrow$  n$D$ transitions. For 
states with $n\approx20$, the transition wavelength $\lambda_{probe}$ is about 510nm. In Fig.~\ref{Fig2} 
we show the basic principle of the experiment that will be analysed.

\begin{figure}[!t]
\includegraphics[width=75mm]{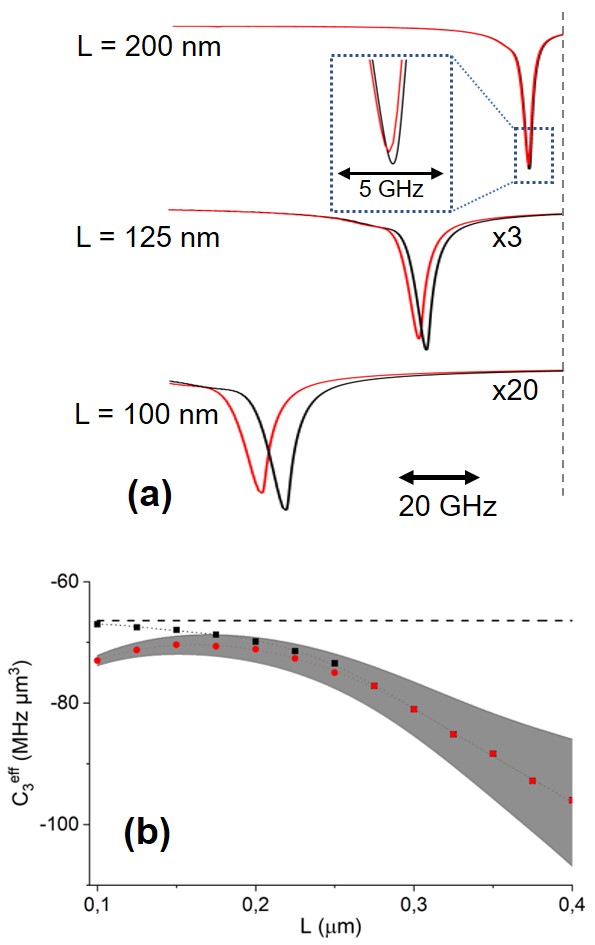}\\
\caption{(a) Thin-cell transmission spectra for three different thicknesses, calculated using 
Eq.~(\ref{eq:S_T}) ($S_T$) with $C_3$=4.15 $\mathrm{MHz} \mathrm{\mu m^{3}}$ and $C_5$=0.45 
$\mathrm{kHz} \mathrm{\mu m^{5}}$, corresponding to Cs 16$S_{1/2}$ atoms. Dashed vertical line: 
position of the transition frequency in the volume. Straight red lines: calculations including both
dipole and quadrupole interactions ($C_3$ and $C_5$). Black lines: dipole interactions only. 
The transmission amplitude decreases with thickness by a factor indicated in the figure.  
(b) Displacement of the transmission dip $C_3^{eff}$ with respect to the volume transition frequency 
(dashed line), multiplied by $L^3$, as a function of cell thickness. Red dots: calculations with full CP 
potential. Black dots: dipole interactions only. Horizontal dashed line: CP dipole shift in the centre 
of the cell multiplied by $L^3$.
\label{Fig3}}
\end{figure}
Inter-atomic collisions and radiation trapping in the atomic vapour redistribute the initially velocity
selected 6$P$ population to many atomic velocities and hyperfine states. This allows us to consider our 
atoms essentially as two level systems in linear interaction with the probe excitation field. Thin 
cells form a low finesse Fabry-Perot cavity that eventually mixes the backward (reflection $I^{lin}_R$) 
and forward (transmission $I^{lin}_T$) response of the polarised atomic vapour \cite{Dutier_josab_2003} given by:
\begin{gather}
I^{lin}_R \propto \frac{N \mu^2}{\mathcal{F}} \int_{0}^{\infty} \frac{dv_z}{v_z}W(v_z)\int_{0}^{L} dz \int_{0}^{z} 
dz' e^{2ikz} e^{\frac{\mathcal{L}(z')-\mathcal{L}(z)}{v_z}} ,\\
I^{lin}_T \propto \frac{N \mu^2}{\mathcal{F}} \int_{0}^{\infty} \frac{dv_z}{v_z}W(v_z)\int_{0}^{L} dz \int_{0}^{z} 
dz' e^{\frac{\mathcal{L}(z')-\mathcal{L}(z)}{v_z}}\,.
\end{gather}
For a symmetric Fabry-Perot cavity with reflection coefficient $r$ for both interfaces the transmission signal ($S_T$) that consists of the homodyne beating between the atomic 
response with that of an empty cavity is given by
\begin{equation}
\label{eq:S_T}
S_T \propto \frac{2 N \mu^2}{|\mathcal{F}|^2}\Re[I^{lin}_T  +r^2 e^{2 i k L}I^{lin}_T -2 r I^{lin}_R].
\end{equation}

Here, $\mathcal{F}=1-r^2 e^{2ikL}$ with $r$ the reflection coefficient of the windows and 
$k=2\pi/\lambda_{probe}$. The atomic velocity along the probe beam propagation axis is denoted as $v_z$, 
the atomic vapour density inside the cell is $N$ and the transition dipole moment is $\mu$. In the above 
equations, the functions $\mathcal{L}(z')- \mathcal{L}(z)$ are
\begin{equation}\label{3.24}
    \mathcal{L}(z')-\mathcal{L}(z)  = \int_{z}^{z'} \left[ \Gamma/2-i(\delta+2 \pi \Delta f_{CP}(\xi)-k v_z) \right] d\xi\,,\nonumber
\end{equation}
where $\Delta f_{CP}(z)$ is the CP shift of the Rydberg state (the shift of the 6$P$ state is 
negligible) inside the thin-cell cavity. The shift can be separated into a dipole component 
and a quadrupole component that depend on the $C_3$ 
and $C_5$ coefficients, respectively. As full calculation of the CP shift inside a cavity requires a 
more elaborate theory, we will not consider resonant effects due to surface polaritons and simply add the potentials of both walls, neglecting the infinite series of multiple images. 
In this case, the dipole and quadrupole shifts in the middle of the cell become $-16C_3/L^3$ and 
$-64C_5/L^5$, respectively. 

In Fig.~\ref{Fig3}(a) we show the transmission spectrum of a resonant 6$\mathrm{P_{1/2}}$ $\rightarrow$ 
16$\mathrm{S_{1/2}}$ beam through a thin cell of three different thicknesses. The atom-surface 
interaction coefficients calculated for the 16$\mathrm{S_{1/2}}$ state of cesium (with a Bohr diameter of $\approx$ 15nm) are $C_3$=4.15 
$\mathrm{MHz} \mathrm{\mu m^{3}}$ and $C_5$=0.45 $\mathrm{kHz} \mathrm{\mu m^{5}}$. The red curves 
represent the calculated spectra using both dipole and quadrupole interactions, whereas for black curves 
the quadrupole interactions are omitted. The effect of quadrupole interactions becomes observable when 
the atomic vapour is confined at thicknesses smaller than 200nm. At L=100nm the additional quadrupole shift 
exceeds the predicted spectral width, suggesting that the $C_5$ 
coefficient could be measurable with a thin-cell spectroscopy experiment. Figure~\ref{Fig3}(b) shows the 
predicted displacement of the transmission dip $C_3^{\rm eff}$ with respect to the volume resonance away 
from the surface, multiplied by $L^3$. When quadrupole interactions are ignored (black dots), 
$C_3^{\rm eff}$ tends towards $16C_3$ (dashed horizontal line), suggesting that when the cell is very thin, 
the dominant spectral contribution derives from atoms at the centre of the cell. For increasing cell 
thicknesses, the contribution of layers closer to the walls becomes more prominent leading to a larger 
$C_3^{\rm eff}$. Red dots represent calculations including both dipole and quadrupole interactions. From 
Fig.~\ref{Fig3}(b) we can see that quadrupole interactions have no visible effect for thicknesses
larger than 250nm. The extent of the gray shaded area is $L^3 w/5$, where $w$ is the width of the 
transmission spectrum. The gray shaded area gives an indication of the capability of the proposed 
experiment to discern between the two models (black and red dots). Note that the discerning 
capability also depends on the signal to noise ratio of the experiment. The reduction in signal 
amplitude is noted in Fig.~\ref{Fig3}(a) next to the transmission curves. The effects of higher-order 
interactions are evident for larger cell thicknesses when probing higher-lying states such as 
$28\mathrm{S_{3/2}}$ via the 6$\mathrm{P_{1/2}}$ $\rightarrow$ 28$\mathrm{S_{3/2}}$ transition (see 
Supplemental Material). 

Our above analysis assumes that Rydberg atoms interact with surfaces only via CP interactions. However, 
Stark shifts due to adsorbed atoms or parasitic electric field in the surface of the dielectric windows 
are known to be an important problem in precision atom-surface experiments \cite{cornell_PRA_2004, 
Cornel_PRA_2004b, Cornel_PRA_2007, Pereira&Ballad_2023}. In particular, high-lying states become 
extremely sensitive to electric fields \cite{Hatermann&Fortagh_PRA_2012, kublernatphot2010} as their 
polarisability scales more rapidly ($\alpha \propto {n^{\star 7}}$) than the $C_3$ coefficient.
Our analysis suggests that higher-order CP effects can be measurable even with relatively low-lying 
Rydberg states with ${n^{\star}}\approx 10\ldots 15$, for which atom-surface interaction experiments 
have already been demonstrated \cite{SandoghdarPRL1992, SandoghdarPRA1996}. This is a powerful 
indication that vapour cell spectroscopic experiments could provide excellent testbeds for further 
exploring Casimir-Polder physics.

\acknowledgments
We acknowledge insightful discussions with Martial Ducloy and discussions with Isabelle Maurin on the numerical modelling of thin cell spectra. This work was financially supported by the ANR-DFG 
grant SQUAT (Grant Nos. ANR-20-CE92-0006-01 and DFG SCHE 612/12-1), the DAAD and Campus France (via the PHC-PROCOPE programme,
grant No. 57513024), and the French Embassy in Germany (via the Campus-France PHC-Procope project 44711VG and via PROCOPE Mobilité, project DEU-22-0004 
LG1).

\bibliography{RydbergtheoryPRR}


\end{document}